\documentclass[aps,preprint,endfloats*,showpacs]{revtex4}

\usepackage{graphicx}
\usepackage{subfigure}
\begin{document}

\title{Sign-symmetry of temperature structure functions}
\author{Konstantinos G.~Aivalis}
\affiliation{Department of Mechanical Engineering, Yale University,
New Haven, CT 06520, USA}
\author{Susan Kurien}
\affiliation{Center for Nonlinear Studies and Theoretical Division,
Los Alamos National Laboratory, NM 87455}
\author{J\"org Schumacher}
\affiliation{Fachbereich Physik, Philipps-Universit\"at, D-35032
Marburg, Germany}
\author{Katepalli R.~Sreenivasan}
\affiliation{International Centre for Theoretical Physics, Strada
Costiera 11, 34014 Trieste, Italy}
\date{\today}

\begin{abstract}
New scalar structure functions with different sign-symmetry properties
are defined. These structure functions possess different scaling
exponents even when their order is the same. Their scaling
properties are investigated for second and third orders, using
data from high-Reynolds-number atmospheric boundary layer. It is
only when structure functions with disparate sign-symmetry
properties are compared can the extended self-similarity detect
two different scaling ranges that may exist, as in the example of
convective turbulence.

\end{abstract}
\maketitle

\section{Introduction}
A problem of broad interest is the advection and diffusion of
passive scalars in a turbulent flow. The classical paradigm of a
passive scalar is the temperature field $\theta({\bf x},t)$ when
the heating is small. The temperature increments $\Delta_{\bf
r}=\theta({\bf x} + {\bf r}) - \theta({\bf r})$ have been studied
in the literature \cite{Obukhov,Corrsin,monin} in search of
scaling in the intermediate range of scales that are unaffected
directly by either the stirring mechanism or diffusive and viscous
effects. For examples of early and recent experimental studies,
see \cite{antonia} and \cite{moisy}, respectively. The following
two types of moments of $\Delta_{\bf r}$, the so-called structure
functions of order $n$, have been employed:
\begin{eqnarray}
   S_n({\bf r}) &=& \langle {(\theta({\bf x} + {\bf r}) - \theta({\bf
   x}))}^n\rangle = \langle \Delta_{\bf r}^n \rangle ,
   \label{str_def}\label{Sn}\\
   S_{|n|}({\bf r}) &=& \langle {|\theta({\bf x} + {\bf r}) - \theta({\bf
   x})|}^n\rangle =\langle |\Delta_{\bf r}|^n \rangle.
   \label{Snabs}
\end{eqnarray}
Here, ${\bf r}$ is the separation vector between two spatial
positions, and $\langle\cdot\rangle$ defines a suitable ensemble
average. For convenience we will call (\ref{Sn}) the {\it normal}
structure functions and (\ref{Snabs}) the {\it absolute} structure
functions. When ${r \equiv |\bf r|}$ is small compared to the
large scale $L$, both structure functions are homogeneous, i.e.,
independent of ${\bf x}$. Clearly, (\ref{Sn}) and (\ref{Snabs})
coincide for even $n$. However, remembering that normal odd
moments are zero in the absence of a mean temperature gradient
while that is not so for absolute odd moments, one may expect, for
odd $n$, that there might be perceptible differences in the two
classes of structure functions. It is useful to quantify these
differences and stress the reasons why they might be important.

\begin{figure}
\centering
\subfigure{\label{fig:fig0a}
\includegraphics[scale=0.35]{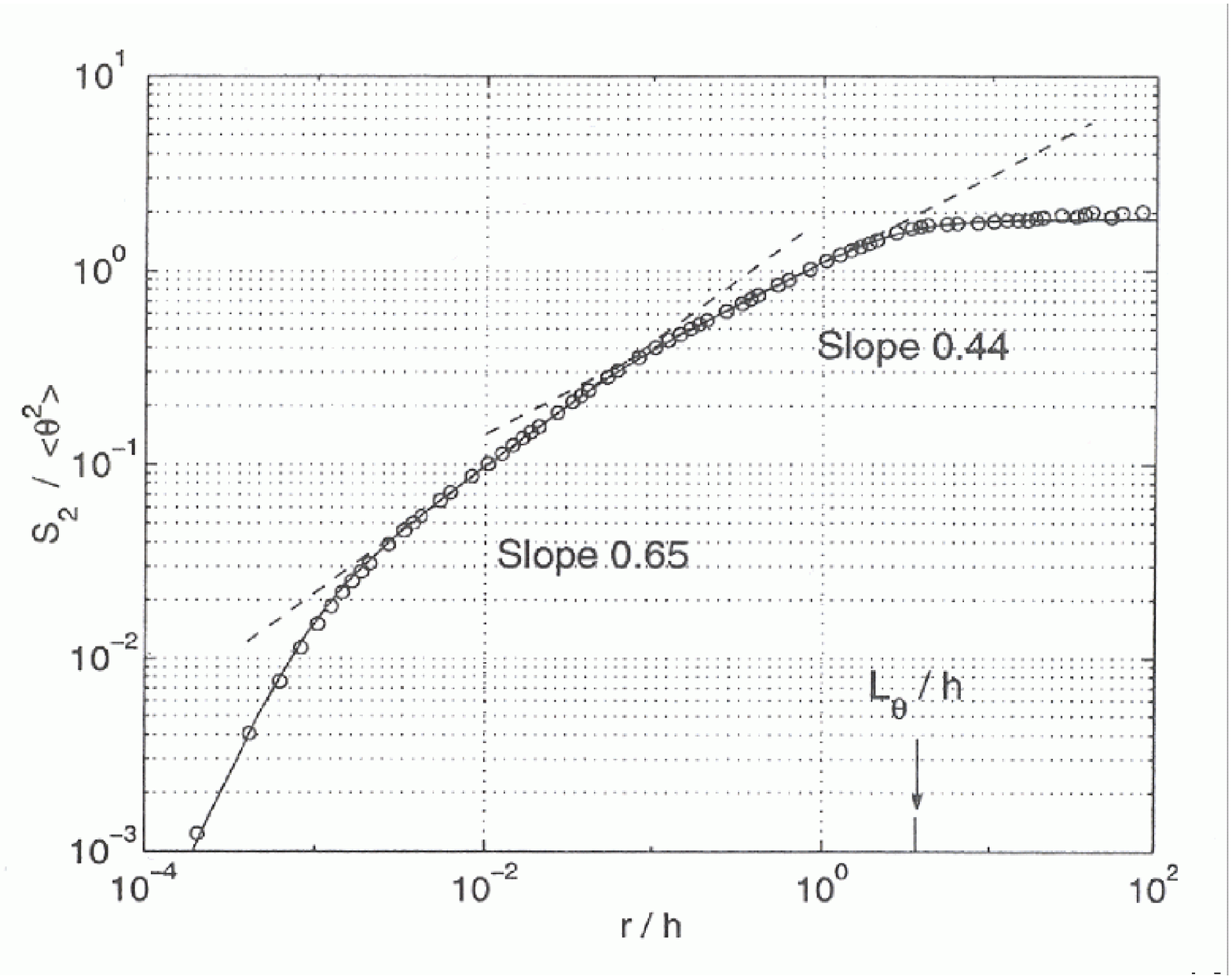}}
\subfigure{\label{fig:fig0b}
\includegraphics[scale=0.35]{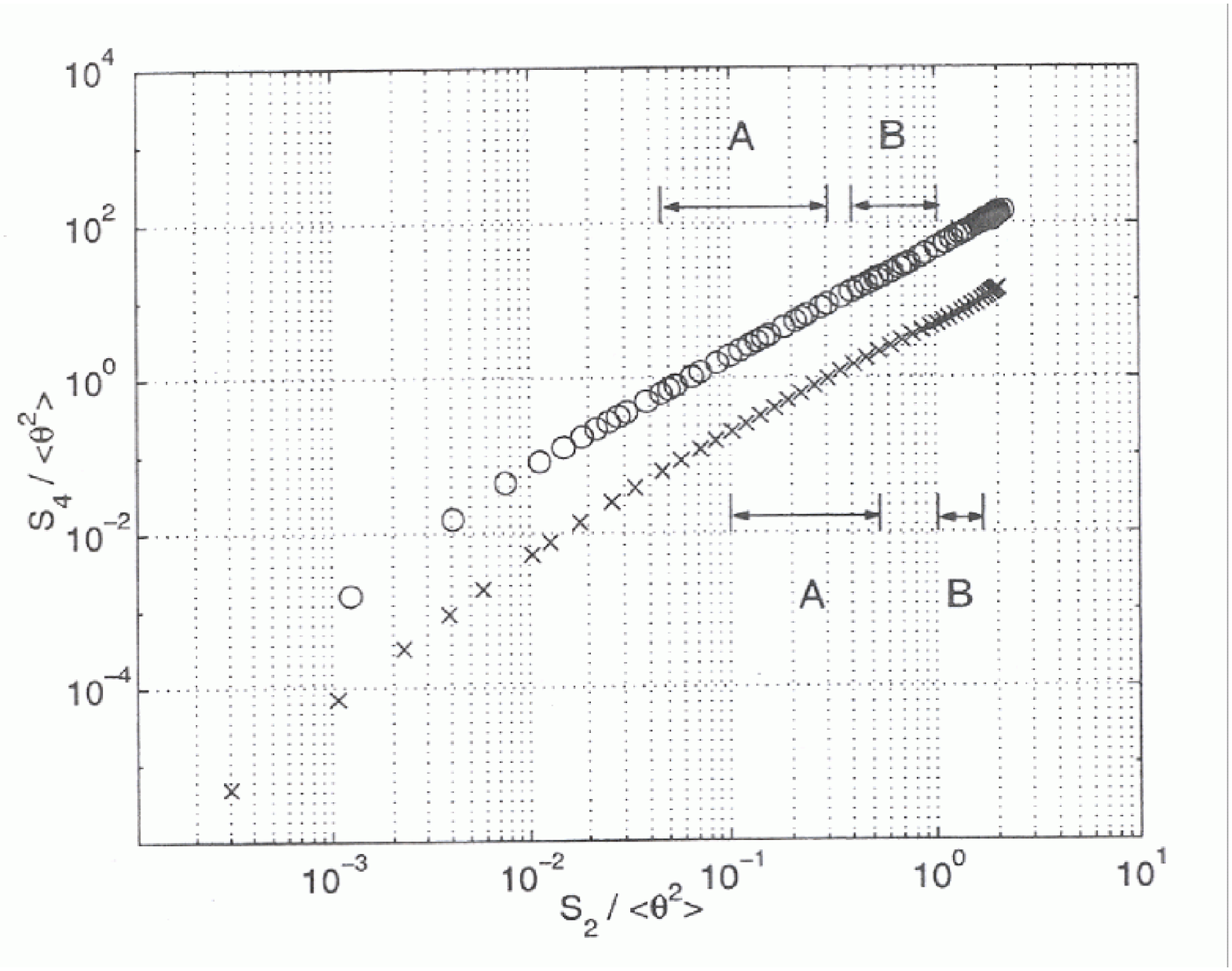}}
\caption{\label{S2} Normal structure function of the temperature
fluctuations. Top figure: $S_2({\bf r})$ is shown for height
$h=1.75$ m above the ground. Shown by the vertical arrow on the
abscissa is the integral scale $L_{\theta}$ for the temperature
fluctuations. Bottom figure: $S_4({\bf r})$ as a function of
$S_2({\bf r})$. The data corresponding to the top figure are marked
as circles (multiplied, for convenience, by a factor of 10). A and
B, which have the same slope in the bottom figure, correspond to
the two scaling ranges with slopes of 0.65 and 0.44, respectively,
in the top figure. The crosses in the bottom figure correspond to
another similar situation where A and B, again, have two
distinctly different slopes for normal structure functions. Both
figures are reproduced from Ref.~\cite{kga_krs_01}.}
\end{figure}

The following two comments put the present work in perspective.
The first concerns the extraction of scaling exponents of
structure functions using the Extended Self-Similarity (ESS)
\cite{Benzi96}. Instead of examining the scaling of normal
structure functions, $S_{n}(r)$, with respect to the scale
separation $r$ directly, the practice is to examine the scaling
relative to another structure function, say $S_{m}(r)$, $m \ne n$.
This usually leads to the extension of a possible algebraic
scaling range, and the relative scaling exponent, $S_{n}(r)\sim
S_{m}(r)^{\zeta_{n,m}}$ where $\zeta_{n,m} = \zeta_n/\zeta_m$, can
be obtained with greater confidence. In the literature, the
implementation of the method has often mixed up normal structure
functions and absolute structure functions without exploring the
differences between them. Further, it was recently shown that in
the presence of convection, ESS fails to show the existence of two
distinct scaling ranges---the nearly passive behavior at small
scales and the dominance of buoyancy at large scales
\cite{kga_krs_01}. This point is illustrated in Fig.~\ref{S2}. In
the top figure, we show the second order structure function with
clearly separated scaling ranges; in the bottom figure, the
corresponding ESS plot is shown to result in a line of nearly
constant slope, without making the necessary distinction between
the regions marked A and B. (The second set of data in the bottom
figure corresponds to another set of measurements, with
qualitatively similar conclusion.)

The second comment is that the structure functions, Eqs.\
(\ref{Sn}) and (\ref{Snabs}), have different sign-symmetry
properties. One has to distinguish the sign-symmetry with respect
to the reversal of increments $\Delta_{\bf r}\to -\Delta_{\bf r}$
from that with respect to spatial reflection (also known as
parity)---that is, for the transformation ${\bf r}\to -{\bf r}$.
The absolute structure functions (\ref{Snabs}) remain unchanged
for all orders $n$ under both transformations; they have {\it
even} sign-symmetry. The same is true for even-order normal
structure functions but the odd-order normal structure functions
change sign under both transformations; they have {\it odd}
sign-symmetry or sign-antisymmetry. Note that $\Delta_{\bf r}\to
-\Delta_{\bf r}$ does not follow from ${\bf r}\to -{\bf r}$
although, for homogeneous turbulence, both transformations have
the same effect on (\ref{Sn}). We wish to comment on ESS in the
light of the sign-symmetry properties of structure functions. For
this purpose, it is convenient to introduce new types of structure
functions which explicitly emphasize the sign-symmetry with
respect to the increment $\Delta_{\bf r}$. This is our basic goal.

These new quantities, along with (\ref{Sn}) and (\ref{Snabs}),
will be calculated from temperature data from a
high-Reynolds-number atmospheric boundary layer. The Taylor
microscale Reynolds number is about 3500. Since the experiments
have been described in some detail in \cite{Kurien1}, we shall
mention only a few details here. Measurements were made in a
boundary layer above salt flats of the Dugway Proving Ground in
Utah, at a height of 1.75 m above the ground. The ground was
smooth on the order of a millimeter. Taylor hypothesis was used
and $u_{rms}/\overline{U}$ was about 7\% with the mean speed
$\overline{U}=0.72$ ms$^{-1}$. Measurements were made at various
times of the day, covering intensely convective motion in late
afternoon, essentially neutral conditions in the evening hours and
somewhat stable conditions until about 11 PM. The data records
chosen for analysis corresponded to constant wind conditions in
magnitude as well as direction. Temperature fluctuations were
measured by two cold wires of 0.6 $\mu$m diameter and 1 mm length.
The data acquisition system was a standard constant-current
anemometer system operated at a small enough current to minimize
velocity sensitivity.

The paper is organized as follows. In Sec.\
\ref{sec:decomposition}, we will construct the new types of
structure functions, in line with the approach of \cite{Sreeni96}
for velocity increments. Since these structure functions are of
intrinsic interest, they are presented first before considering
ESS. Section \ref{sec:ess} contains the results of the ESS data
analysis for structure functions with even and odd sign-symmetry.
In Sec.\ \ref{sec:explanation}, we discuss a possible explanation
of the effects observed in terms of the SO(3) rotation group
decomposition of structure functions of different sign-symmetry.
Some concluding remarks are presented in Sec.\
\ref{sec:conclusions}.

\section{Sign-symmetry with respect to the temperature increment}
\label{sec:decomposition}
Let the probability density function (PDF) of the temperature
increment at fixed distance vector ${\bf r}$ be given by $f
(\Delta_{\bf r})$. First, we can define (\ref{Sn}) and
(\ref{Snabs}) as
\begin{eqnarray}
   S_n&=& \int_{-\infty}^{\infty} \Delta_{\bf r}^n f (\Delta_{\bf r})
d\Delta_{\bf r}
   \label{f_mom}\\
   S_{|n|} &=& \int_{-\infty}^{\infty} {|\Delta_{\bf r}|}^n f (\Delta_{\bf r})
d\Delta_{\bf r}\,.
   \label{fabs_mom}
\end{eqnarray}
The PDF can be decomposed into its symmetric and antisymmetric
parts with respect to the increment as
\begin{eqnarray}
   f_s (\Delta_{\bf r}) &=& \frac{f (\Delta_{\bf r})+ f (-\Delta_{\bf r})}{2}
   \label{sym_def}\\
   f_a (\Delta_{\bf r}) &=& \frac{f (\Delta_{\bf r})- f (-\Delta_{\bf r})}{2}\,.
   \label{ant_def}
\end{eqnarray}
Note that $f_a$ does not have the positive-definite property of a
PDF.

We can also define the positive and negative parts of the PDF as
\begin{eqnarray}
   p (\Delta_{\bf r},\Delta_{\bf r} \geq 0) &=& f (\Delta_{\bf r})
   \label{pos_def}\\
   n (\Delta_{\bf r},\Delta_{\bf r} \geq 0) &=& f (-\Delta_{\bf r}),
   \label{neg_def}\
\end{eqnarray}
and define moments of $\Delta_{\bf r}$ with respect to $f_s$,
$f_a$, $p$ and $n$, respectively, by the following relations:
\begin{eqnarray}
   P_n &=& \int_{0}^{\infty} \Delta_{\bf r}^n ~p(\Delta_{\bf r})
d\Delta_{\bf r},
   \label{P_mom}\\
   N_n &=& \int_{0}^{\infty} \Delta_{\bf r}^n ~n(\Delta_{\bf r})
d\Delta_{\bf r},
   \label{N_mom}\\
   S_{n,s} &=& 2\int_{0}^{\infty} \Delta_{\bf r}^n ~f_s(\Delta_{\bf
     r})
d\Delta_{\bf r},
   \label{fs_mom}\\
   S_{n,a} &=& 2\int_{0}^{\infty} \Delta_{\bf r}^n ~f_a (\Delta_{\bf
     r})
d\Delta_{\bf r}\,.
   \label{fa_mom}
\end{eqnarray}
The following relations are now valid:
\begin{eqnarray}
   S_{n,s} &=& P_n + N_n = S_{|n|}
   \label{evenodd1}\\
   S_{n,a} &=& P_n - N_n\,.
   \label{evenodd2}
\end{eqnarray}
For odd values of $n = 2k+1$, Eq.\ (\ref{evenodd2}) reduces to the
normal odd-order structure function
\begin{equation}
   S_{2k+1,a} = P_{2k+1} - N_{2k+1} = S_{2k+1}\,,
   \label{odd1}
\end{equation}
whereas, for even $n = 2k$, Eq.\ (\ref{evenodd2}) is a new
structure function which is $even$ order, but
sign-$antisymmetric$. It is not possible to have this combination
in terms of either normal or absolute structure functions.

On the basis of the discussion of the sign-symmetries of $f_s$ and
$f_a$, using the definitions (\ref{pos_def}) and (\ref{neg_def}),
it is clear that
\begin{eqnarray}
   P_n + N_n &=& \int_0^{\infty}\Delta_{\bf r}^n (p (\Delta_{\bf r})
+n (\Delta_{\bf r})) d\Delta_{\bf r} \nonumber\\
&=& \int_0^{\infty}\Delta_{\bf r}^n (f (\Delta_{\bf r})
+f (-\Delta_{\bf r})) d\Delta_{\bf r}
   \label{parity1}
\end{eqnarray}
is sign-symmetric, while
\begin{eqnarray}
   P_n - N_n &=&\int_0^{\infty}\Delta_{\bf r}^n(p (\Delta_{\bf r})
-n (\Delta_{\bf r})) d\Delta_{\bf r} \nonumber\\
&=& \int_0^{\infty}\Delta_{\bf r}^n(f (\Delta_{\bf r})
-f (-\Delta_{\bf r})) d\Delta_{\bf r}
   \label{parity2}
\end{eqnarray}
is sign-antisymmetric.

\begin{figure}[t]
\subfigure[Logarithmic local slopes of the various structure
functions for
$n=2$]{\label{fig:fig1a}\centering\includegraphics[scale=0.4]{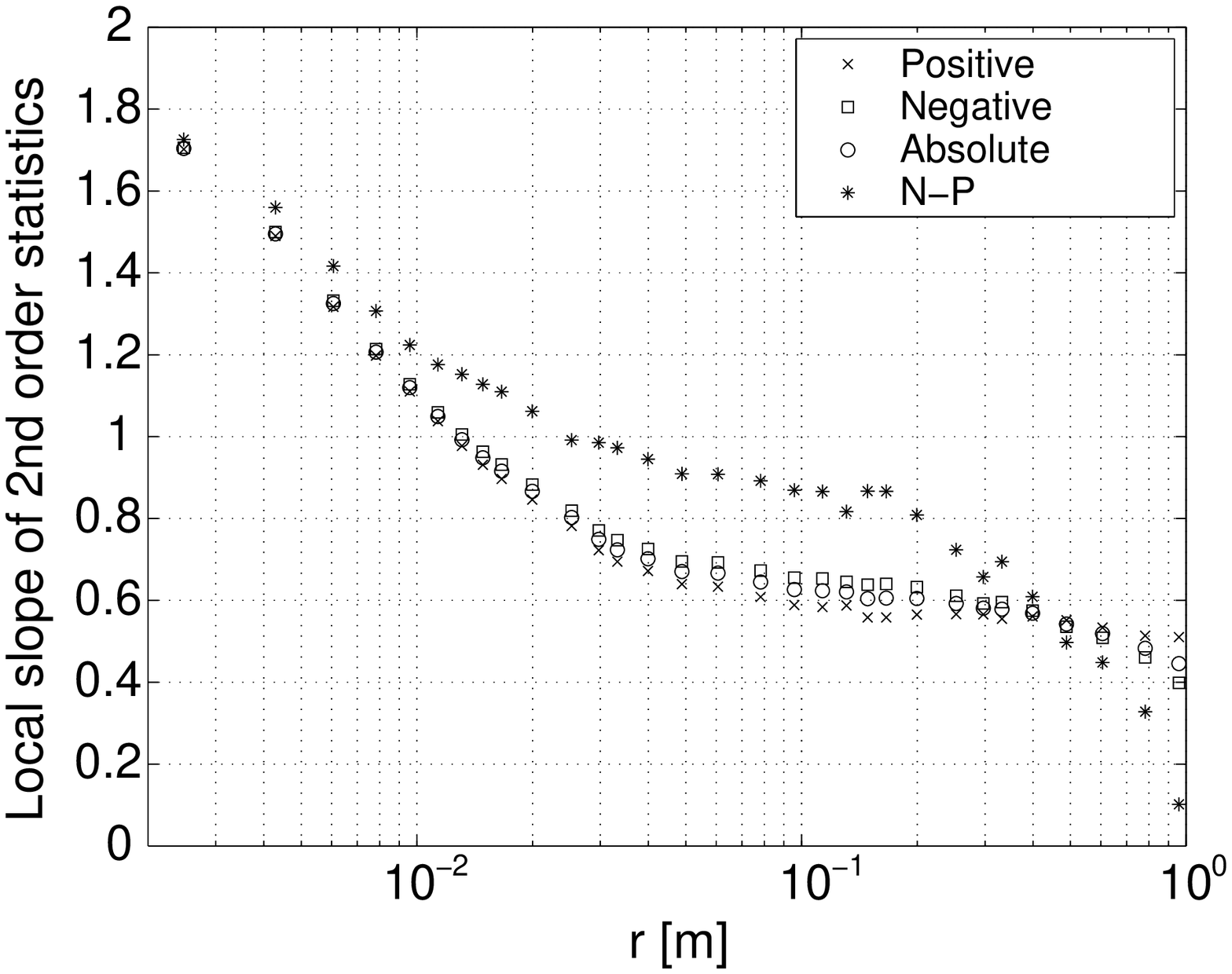}}
\subfigure[Logarithmic local slopes of the various structure
functions for
$n=3$.]{\label{fig:fig1b}\includegraphics[scale=0.4]{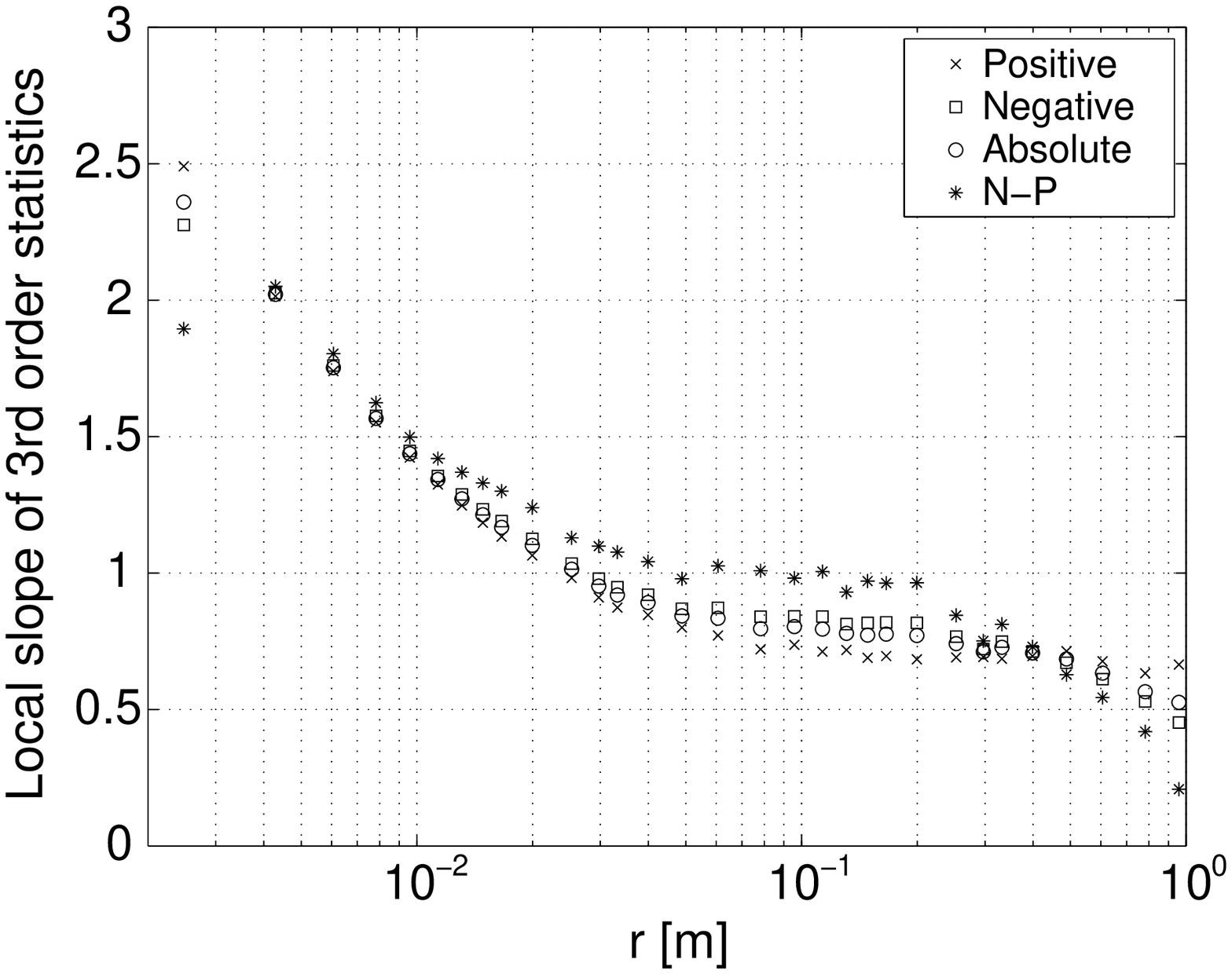}}
\caption{Local slopes in the log-log graphs of four types of
structure functions: positive ($P_n$) and negative ($N_n$) parts,
the sum of positive and negative parts (same as the absolute
moment), $S_{|n|}=P_n+N_n=S_{n,s}$, and the negative minus the
positive part, $|S_n|=N_n-P_n=|S_{n,a}|$.
\label{fig:fig1}}
\end{figure}

To illustrate the differences between normal and absolute
structure functions, we plot in Fig.~\ref{fig:fig1} the
logarithmic local slopes for four types of moments: $P_n$, $N_n$,
$S_{n,s}$ and $S_{n,a}$. Figure~\ref{fig:fig1a} show these
quantities for $n=2$. Here $S_{2,a}$ is the normal (and absolute)
second order structure function; $|S_{2,a}| = N_2 - P_2$ is the
newly defined sign-antisymmetric second order structure function.
Figure \ref{fig:fig1b} shows the same four quantities for $n=3$.
Although the scaling is not impeccable even at this high Reynolds
number (see Ref.\ \cite{Sreeni98} for comments in this regard on
the scaling of velocity structure functions), a scaling tendency
can be discerned in the range between a few mm and a few cm.

Independent of this detail, it is clear that the exponents, if one
were to assign nominal values in the scaling range, are not the
same for all the different structure functions of the same order.
In particular, the second-order sign-antisymmetric structure
function $|S_{2,a}| = N_2 - P_2$ has a substantially larger
exponent than the classical exponent of 2/3. While $S_{3,a} = S_3$
has a scaling exponent close to the Kolmogorov prediction of 1,
the other curves have measurably smaller scaling exponents. In
particular, the absolute structure functions have smaller scaling
exponents than the normal structure function.

At the least, these features are disconcerting to anyone
interested in scaling exponents. An understanding of differences
in the exponents of the various types of structure functions may
help in this regard. This will be attempted in the next section.

\section{Analysis using Extended Self-Similarity}\label{sec:ess}
\begin{figure}[t]
\centering \subfigure[Loglog plot of $S_2$ vs.
$S_{|3|}$]{\label{fig:fig2a}\includegraphics[scale=0.4]{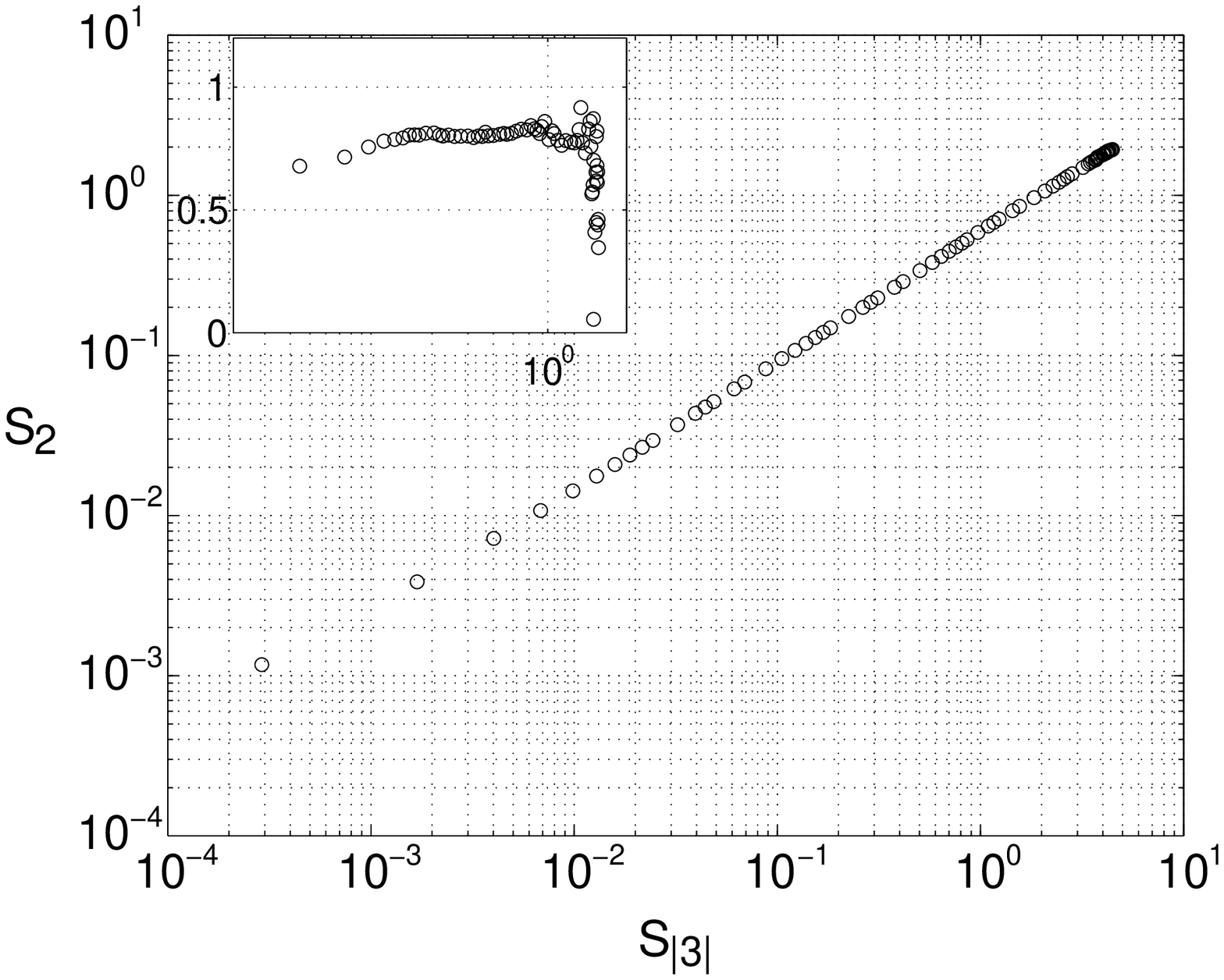}}
\subfigure[Loglog plot of $S_{2,a}$ vs.
$S_{3,a}$]{\label{fig:fig2b}\includegraphics[scale=0.4]{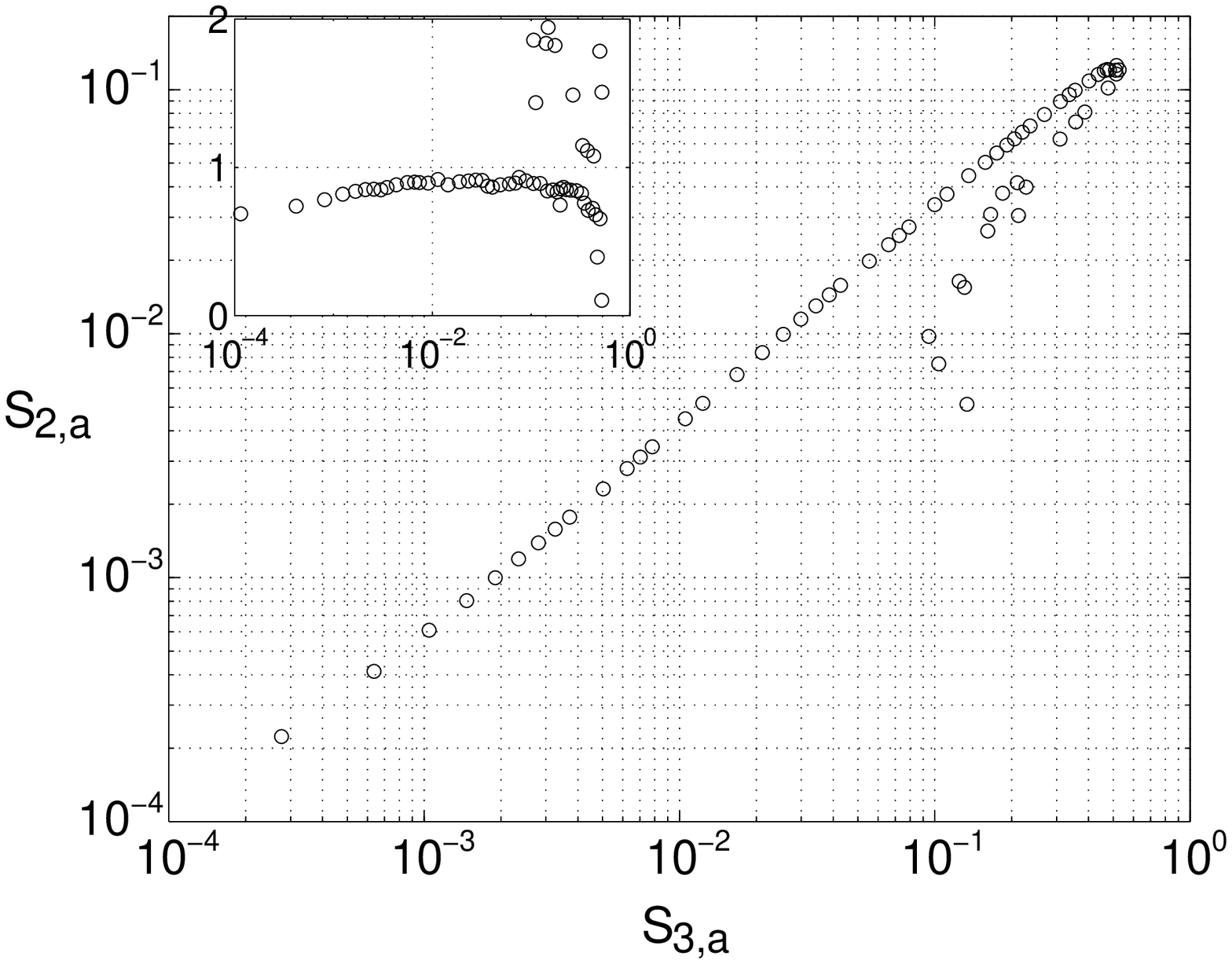}}
\caption{ESS comparisons of structure functions of different order
and same sign-symmetry. Insets show the (logarithmic) local slope
in these ESS coordinates. In this comparison, only one scaling
exponent is seen past the dissipation range. The conditions for
this and other figures to follow are the same as for Fig.\
1.
\label{Sns_vs_Sms}}
\end{figure}
In this section, we shall examine ESS of scalar statistics in the
light of the new objects defined in terms of their parity.  We
first apply ESS as is normally done for velocity statistics: plot
structure functions of all orders against the absolute third-order
structure function. We shall follow this practice for illustrative
purposes, even though the third-order does not have a comparably
significant meaning for temperature. In Fig.~\ref{fig:fig2a}, the
ESS plot of $S_2(r)$ versus $S_{|3|}$ shows a single relative
scaling exponent outside the dissipative range. The inset shows the
local logarithmic slope of the extended self-similarity plot which
is calculated
via $D_{2,|3|}=\mbox{d} \log(S_{2}(r))/\mbox{d} \log(S_{|3|}(r))$.
However, this same
signal has two scaling regions with distinct exponents when
plotted against $r$ (corresponding to the passive range at small
scales and the convective range at large scales, see top panel of Fig.~1). 
This two-exponent scaling behavior is
masked by ESS. This is essentially so even if we plot the
sign-antisymmetric structure function of the second-order against
$S_3$ (which is also sign-antisymmetric, see Sec.\ 2), as seen in
Fig.\ \ref{fig:fig2b}.

\begin{figure}
\centering \subfigure[Loglog plot of $S_3$ vs.
$S_{|3|}$]{\label{fig:fig3a}\includegraphics[scale=0.4]{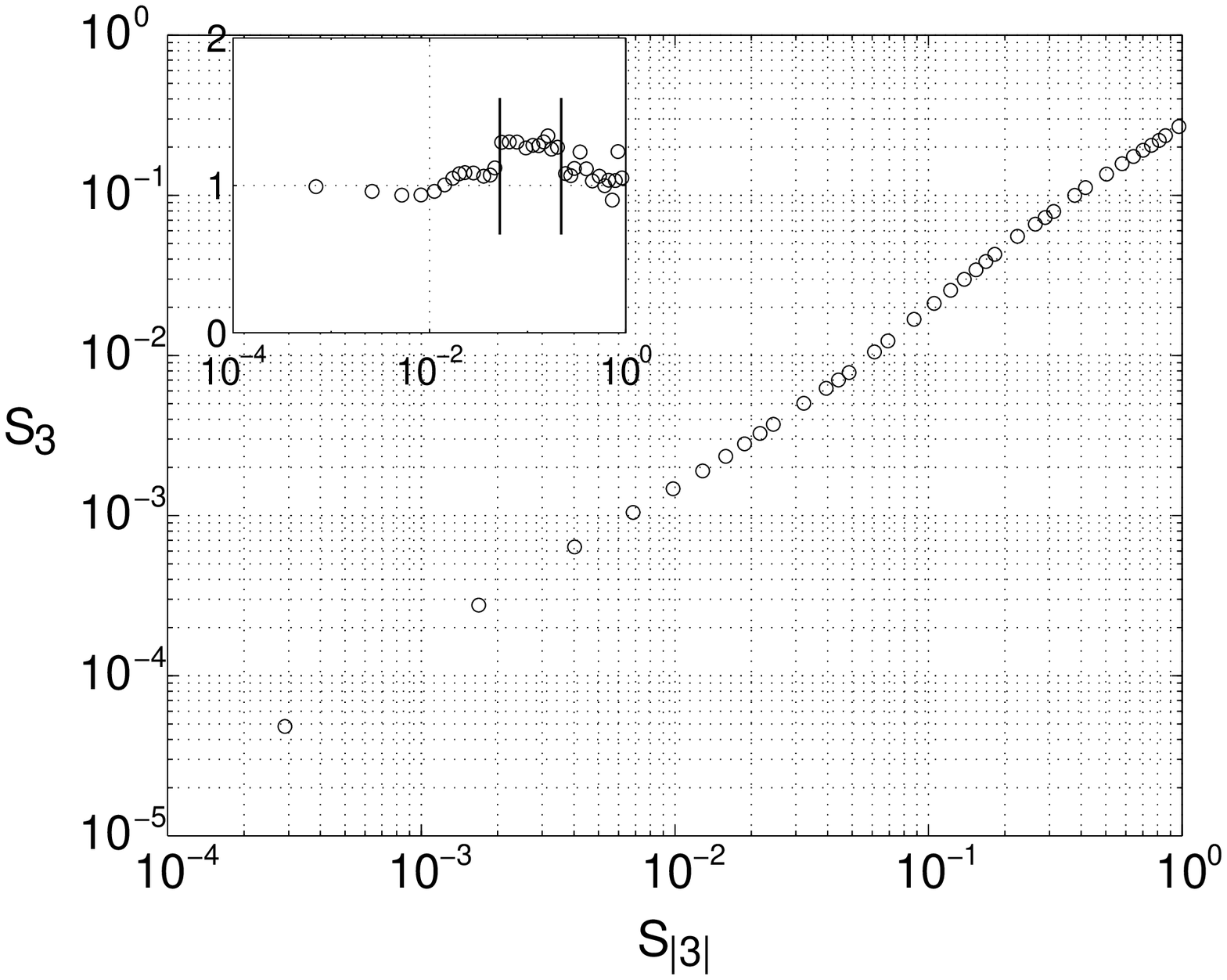}}
\subfigure[Loglog plot of $S_{2,a}$ vs.
$S_{2,s}$]{\label{fig:fig3b}\includegraphics[scale=0.4]{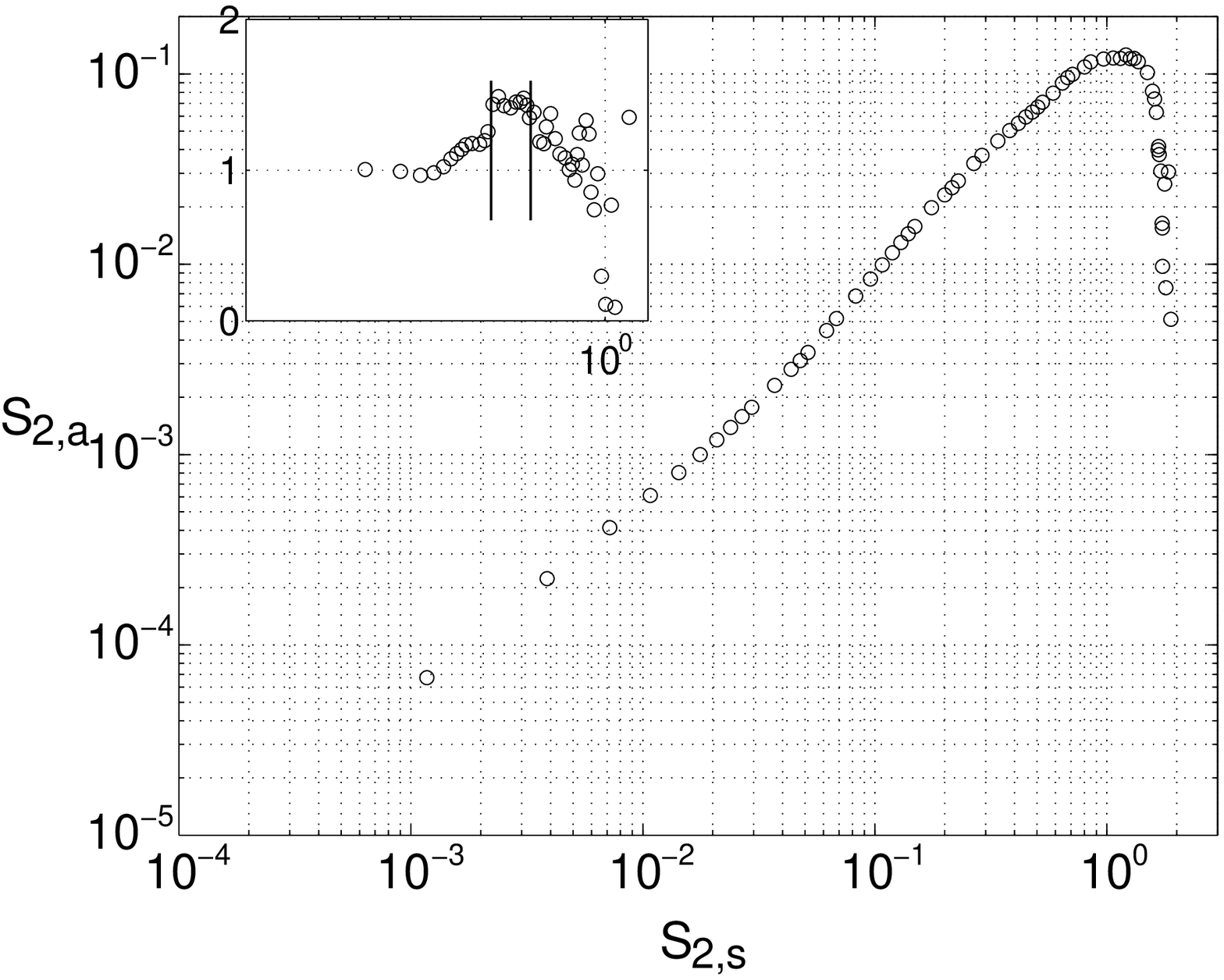}}
\caption{ESS comparisons of structure functions of the same order
and different sign-symmetry. Inset shows the (logarithmic) local
slope in these ESS coordinates. For this comparison, the two
different scaling regimes are retreived. Vertical bars are
inserted in order to show regions of constant local slope.
\label{Sna_vs_Sns}}
\end{figure}

\begin{figure}
\centering \subfigure[Loglog plot of $S_{2,s}$ vs.
$S_{3,a}$]{\label{fig:fig4a}\includegraphics[scale=0.40]{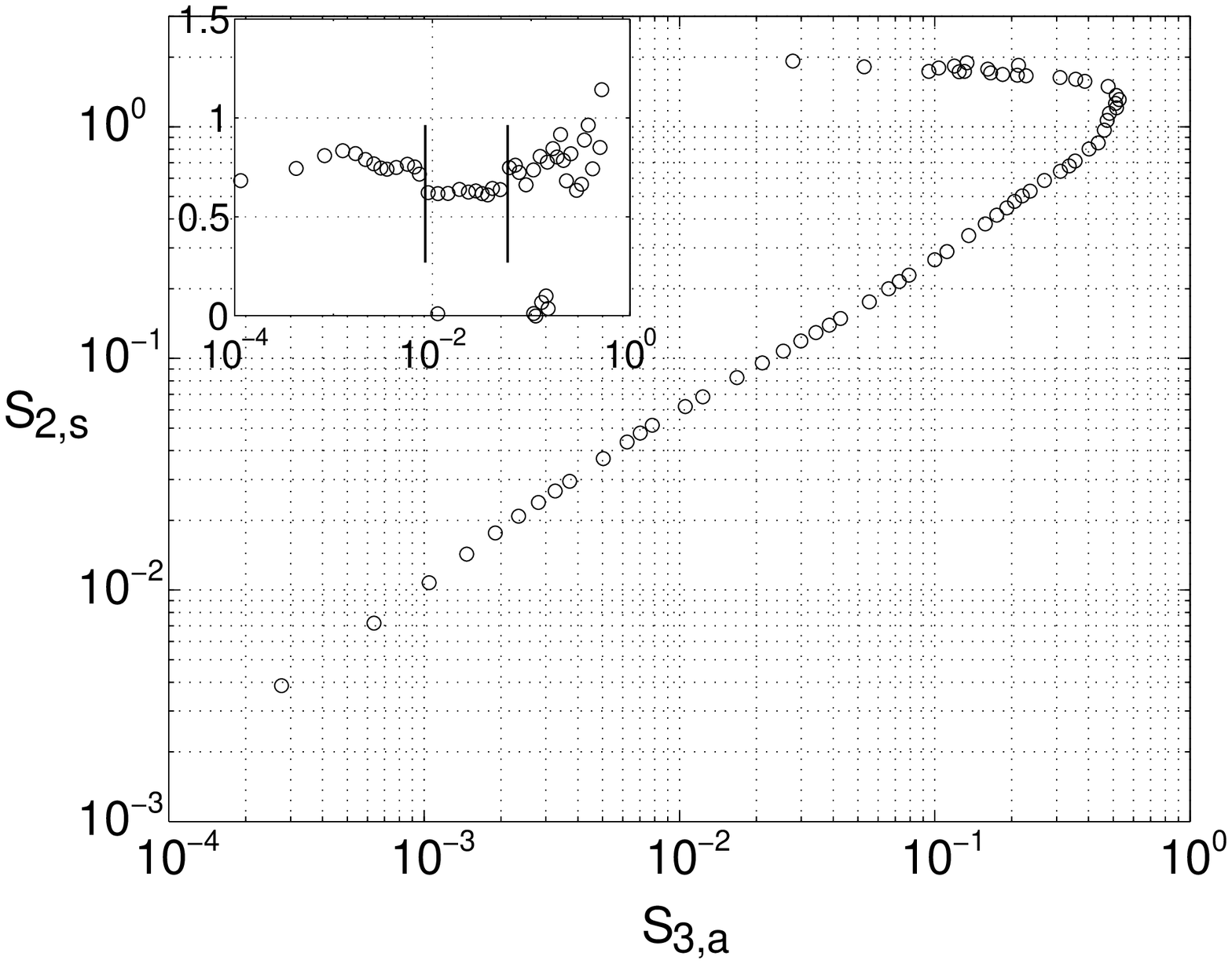}}
\subfigure[Loglog plot of $S_{2,a}$ vs.
$S_{3,s}$]{\label{fig:fig4b}\includegraphics[scale=0.4]{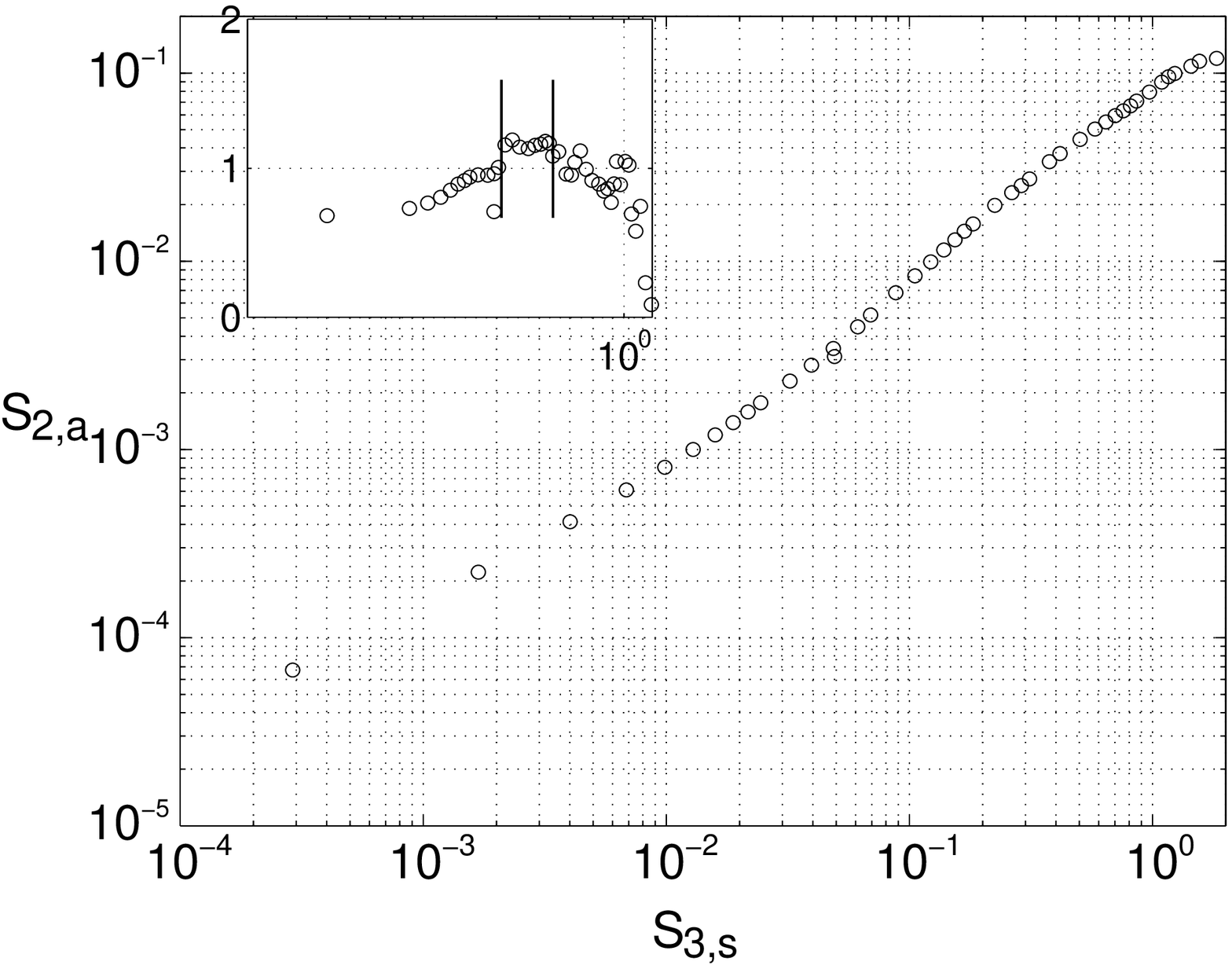}}
\caption{ESS comparisons of structure functions of different order
and different sign-symmetry. The insets show\ the (logarithmic)
local slope in these ESS coordinates. Once again, two different
scaling regimes are retrieved. Vertical bars are inserted in order
to show regions of constant local slope. \label{Sna_vs_Sms}}
\end{figure}

Let us now compare objects of the same order, but different
sign-symmetry.  In Fig.~\ref{fig:fig3a}, we show a plot of
$S_3(r)$ which is odd-parity (made of the antisymmetric
combination of the PDF functions) versus $S_{|3|}(r)$, which is
the corresponding even-parity object of the same order. The two
objects scale identically in the smallest (dissipative/diffusive)
scales with relative exponent of unity, transitioning into a
region of relative exponent 1.27. This corresponds to the inertial
range scaling, as we already know from Fig.~\ref{fig:fig1}. Past
this range, the relative scaling exponent drops back to $\sim 1$.
In Fig.~\ref{fig:fig3b}, we repeat the ESS comparison for the
second-order. Instead of plotting the second-order object against
another order, we choose to compare the even- and odd-parity
manifestations $S_{2,s}(r) = S_2(r)$ and $S_{2,a}(r)$,
respectively. Once again, it is seen that there are two regions of
scaling beyond the dissipative range. The small-scale range has a
relative scaling exponent of about 1 while the large-scale range
has a slope of 1.42. These exponents are as expected from the
direct scaling analysis of \cite{kga_krs_01}.

Finally, when we compare objects of different order $and$
different parity as shown in Fig.~\ref{Sna_vs_Sms}, the dual
scaling range feature persists. Figure~\ref{fig:fig4a} shows
$S_{2,s}=S_2(r)$ versus $S_{3,a}=S_3(r)$ while
Fig.~\ref{fig:fig4b} shows $S_{2,a}$ vs. $S_{3,s}=S_{|3|}(r)$.
Again, we recover two separate scaling ranges, one inertial and
one convective.

From these examples, we can now make the following general
statement. In a process with two distinct scaling ranges, ESS
cannot distinguish between them whenever a comparison is made
between two like-parity objects. However, if ESS compares objects
of $any$ order, but of opposite parity, a second scaling (if it
exists) can be recovered.  This is our major qualitative conclusion.
This kind of comparison has been made
possible by the introduction of the odd-symmetry, even-order objects
$S_{n,a}(r)$ ($n$ even)---which, to our knowledge, have not been
considered before.

\section{Parity of the new moments from the viewpoint of
SO(3) decomposition}\label{sec:explanation}

We shall now discuss the above property of ESS from another
perspective that contains the correct description of the parity
property. We consider an analytical expression for structure
functions with two distinct scaling ranges, separated by a
crossover scale $l_c$. Since we do not know the proper analytical
expression for the structure functions, we use an empirical
interpolation formula. Batchelor's attempt \cite{Batchelor51} in
this direction has been extended variously, in particular in
Refs.\ \cite{stolo_sreeni1,stolo_sreeni2}, for structure functions
of all orders.  Here, we shall extend the specific form proposed
in Ref.\ \cite{Fujisaka01} to the $n$th order as
\begin{eqnarray}
S_{n}(r)\sim \left(\frac{r}{l_c}\right)^{\alpha_{n}}\,
\left[g\left(\frac{r}{l_c}\right)\right]^{\alpha_{n}-\beta_{n}}\,,
\label{Bat1}
\end{eqnarray}
with the function
\begin{eqnarray}
g(x)=\frac{1}{(1+x^{\kappa})^{1/\kappa}}\,, \label{Bat2}
\end{eqnarray}
where the assumed scaling exponents $\alpha_n$ hold for the
passive range, $r\ll l_c$, and $\beta_n$ for the convective range,
$r\gg l_c$, and $\alpha_n > \beta_n$; $g(x)$ is a dimensionless
function that is monotonically decreasing, and the exponent
$\kappa>1$ determines the width of the crossover from one scaling
range to the other.

We make use of the $spatial$ $parity$ of the signed structure
functions by noting that, in homogeneous turbulence, $S_{n,s}({\bf
r}) = S_{|n|}({\bf r})$ is even-parity with respect to reflection
under ${\bf r} \rightarrow -{\bf r}$ for all $n$; the odd-order
normal structure functions $S_{2k+1,a}({\bf r}) = S_{2k+1}({\bf
r})$ are odd-parity; the newly defined $S_{2k,a}$, even-order
sign-$antisymmetric$ statistics are odd-parity as well. So, in the
homogeneous case the sign-symmetry with respect to the increment
is the $same$ as the sign-symmetry with respect to the spatial
orientation of ${\bf r}$ (i.e.\ parity) and this allows for the
following decomposition of the objects defined in Sec.\ II.

The recently developed SO(3) group decomposition \cite{Arad99}
applies conveniently to the symmetric and antisymmetric structure
functions. For the scalar case, the basis functions are spherical
harmonics $Y_{l,m}({\bf \hat r})$ \cite{Kurien2}. The even or odd
{\em spatial} parity is carried by the angular dependence in the
spherical harmonics, in effect by the index $l$, as
\begin{equation}
Y_{l,m}(-\hat{\bf r})= (-1)^l Y_{l,m}(\hat{\bf r})
\end{equation}
Perfectly isotropic objects would contain an $l=0$ sector only,
i.e., no angular dependence would be present. Sign-symmetric
functions are composed of only even sectors while
sign-antisymmetric functions are composed only of odd sectors.
Thus,
\begin{eqnarray}
&S_{n,s}(r,\theta,\phi)&=S_{n,s}^{(0)}(r)+S_{n,s}^{(2)}(r,\theta,\phi)+\dots
,\\
&=&\sum_{k=0}^{\infty}\sum_{m=-2k}^{2k}A_{2k,m}r^{\zeta_n^{(2k)}}
                         Y_{2k,m}(\theta,\phi),\nonumber
\label{symm1}
\end{eqnarray}
where the superscripts $(0)$, $(2)$, $\dots$ denote the
even-parity contributions allowed, and
\begin{eqnarray}
&S_{n,a}(r,\theta,\phi)&=S_{n,a}^{(1)}(r,\theta,\phi)+
                         S_{n,a}^{(3)}(r,\theta,\phi)+\dots
,\\
&=&\sum_{k=0}^{\infty}\sum_{m=-2k-1}^{2k+1}A_{2k+1,m}
r^{\zeta_n^{(2k+1)}} Y_{2k+1,m}(\theta,\phi),\nonumber
\label{symm2}
\end{eqnarray}
where the superscripts $(1)$, $(3)$, $\dots$ denote the odd-parity
contributions allowed. We can substitute the algebraic scaling
form of (\ref{Bat1}) for each scaling term in (\ref{symm1}) and
(\ref{symm2}) to obtain
\begin{eqnarray}
S_{n,s}(r,\theta,\phi)=\sum_{k,m}&& A_{2k,m}
                       \left(\frac{r}{l_c}\right)^{\alpha_{n}^{(2k)}}\nonumber\\
&\times&\left[g\left(\frac{r}{l_c}\right)\right]^
{\alpha_{n}^{(2k)}-\beta_{n}^{(2k)}} \nonumber\\
&\times& Y_{2k,m}(\theta,\phi)\,, \label{Bat3}
\end{eqnarray}
and
\begin{eqnarray}
S_{n,a}(r,\theta,\phi)=\sum_{k,m}&&A_{2k+1,m}
                       \left(\frac{r}{l_c}\right)^
                       {\alpha_{n}^{(2k+1)}}\nonumber\\
&\times&\left[g\left(\frac{r}{l_c}\right)\right]^
{\alpha_{n}^{(2k+1)}-\beta_{n}^{(2k+1)}} \nonumber\\
&\times& Y_{2k+1,m}(\theta,\phi)\,. \label{Bat3a}
\end{eqnarray}

The finding that ESS yields the same relative exponents, when
comparisons are made of even-parity statistics, means that for all
sectors and for any pair of orders $(p,q)$ we have
\begin{equation}
\frac{\alpha_{p}^{(2k)}}{\alpha_{q}^{(2k)}}=
\frac{\beta_{p}^{(2k)}}{\beta_{q}^{(2k)}}\,. \label{even_const}
\end{equation}
Equivalently, for $p<q$, it follows that
\begin{equation}
\frac{\alpha_{p}^{(2k)}}{\alpha_{p}^{(2k)}\,\Pi_{i=p}^{q-1}
\left(1+\delta_i^{(2k)}\right)}=
\frac{\beta_{p}^{(2k)}}{\beta_{p}^{(2k)}\,\Pi_{i=p}^{q-1}
\left(1+\delta_i^{(2k)}\right)}\,, \label{Bat4}
\end{equation}
where the quantities
\begin{equation}
\delta_p^{(2k)}=\frac{\alpha_{p+1}^{(2k)}-\alpha_p^{(2k)}}{\alpha_p^{(2k)}}\,,
\label{odd_const}
\end{equation}
are the relative increments between successive orders. The
constraint (\ref{even_const}) means that the relative increments
in both convex functions, $\alpha_{p}(p)$ as well as
$\beta_{p}(p)$, are equal. This was already shown in Ref.\
\cite{Benzi96}, where physically different turbulent states such
as the Kolmogorov turbulence in three dimensions, thermal
convection as well as magnetohydrodynamic turbulence showed the
same relative ESS exponents. We suppose, therefore, that as long
as the different regimes belong to the same ``universality
class"---i.e., if they have the same relative increments
$\delta_n^{(2k)}$ for all orders $n$---ESS will simply mask the
presence of distinct scaling regimes.

The argument for the odd-parity ESS comparisons is similar, now
with the constraint
\begin{equation}
\frac{\alpha_{p}^{(2k+1)}}{\alpha_{q}^{(2k+1)}}=
\frac{\beta_{p}^{(2k+1)}}{\beta_{q}^{(2k+1)}}\,.
\end{equation}
One remains within a universality class when plotting $S_{p,a}(r)$
as functions of $S_{q,a}(r)$, but it can be expected that the
relative exponents will be different from those for the even
parity case.

We now turn to the case where $S_{p,s}(r)$ are compared to
$S_{q,a}(r)$, which are objects with different symmetries. When
proceeding as in (\ref{even_const}), (\ref{Bat4}), and
(\ref{odd_const}) we obtain
\begin{equation}
\frac{\alpha_{p}^{(2k)}}{\alpha_{p}^{(2k+1)}\Pi_{i=p}^{q-1}
\left(1+\epsilon_i^{(2k+1)}\right)}=
\frac{\beta_{p}^{(2k)}}{\beta_{p}^{(2k+1)}\Pi_{i=p}^{q-1}
\left(1+\epsilon_i^{(2k+1)}\right)}.\nonumber\label{Bat5}
\end{equation}
For insensitivity to different scaling regimes, the following must
hold true:
\begin{equation}
\frac{\alpha_{p}^{(2k)}}{\alpha_{p}^{(2k+1)}}=
\frac{\beta_{p}^{(2k)}}{\beta_{p}^{(2k+1)}}\,. \label{Bat6}
\end{equation}
Recall that the superscript $(2k)$ corresponds to even parity
objects while $(2k+1)$ corresponds to odd parity. We expect that
several mechanisms, such as the breaking of reflection symmetry by
a mean gradient for scales $r<l_c$ or by buoyancy effects for
scales $r>l_c$, will lead to different ratios on the left and
right hand side of the Eq.~(\ref{Bat6}), respectively, thus
leading to a ``discontinuity'' in scaling. While this discussion
cannot be a proof of our main observation, it shows that the two
scaling regimes belong to different universality classes when
different parities are involved. Further, it provides hints for
what kind of quantities have to be investigated in order to shed
more light on the issue, in experiments as well as in numerical
simulations.

\section{Conclusions}\label{sec:conclusions}
We have analyzed moments of temperature increments with respect to
their sign-symmetry properties, and defined sign-symmetric and
sign-antisymmetric components for both even and odd-order
structure functions. ESS analysis of all combinations of symmetry
and order of structure functions indicates that this technique
masks the convective scaling regime. Only if objects of opposite
parity are compared can one recover the distinct scalings.

We have presented a model for how such scaling regimes behave in
the ESS analysis using even and odd orders of a spherical harmonic
expansion. Unfortunately, this model cannot be taken to its
logical conclusion because the various numerical coefficients
cannot be obtained with any certainty from the existing data.
Furthermore, a dependence on the strength of the mean temperature
gradient might modify some of our results as well---e.g.,
differences in the exponents of even order moments in comparison
to odd order moments.

One purpose here has been to point out the possible pitfalls in
using ESS without proper consideration of the symmetry properties
of the statistics being compared. This observation does not
detract from the merits of the method. Indeed, ESS has proved to
be a very useful tool in extracting exponents when the scaling
range with respect to $r$ is short.

\acknowledgments We would like to acknowledge fruitful discussions
with A. Bershadskii. One of us (J.S.) acknowledges support by the
Deutsche Forschungsgemeinschaft, the Feodor-Lynen Fellowship
Program of the Alexander von Humboldt-Foundation and Yale
University. He also wishes to thank the Institute for Physical
Science and Technology at the University of Maryland, and the
International Centre for Theoretical Physics at Trieste for
hospitality.


\begin{thebibliography}{99}
\bibitem{Obukhov} A. M. Obukhov,
Isv. Geogr. Geophys. Ser. {\bf 13}, 58 (1949).

\bibitem{Corrsin}
S. Corrsin,
J. Appl. Phys. {\bf 22}, 469 (1951).

\bibitem{monin}
A. S. Monin and A. M. Yaglom,
{\em Statistical Fluid Mechanics}. MIT Press, Cambridge, MA, 1975.

\bibitem{antonia}
R. A. Antonia, Hopfinger, E., Gagne, Y. and F. Anselmet, Phys.\
Rev.\ A {\bf 30}, 2704 (1984).

\bibitem{moisy}
F. Moisy, H. Willaime, J. S. Andersen and P. Tabeling, Phys.\
Rev.\ Lett.\ {\bf 86}, 4827 (1999).

\bibitem{Benzi96}
R. Benzi, L. Biferale, S. Ciliberto, M. V. Struglia and R.
Tripiccione, Physica D {\bf 96}, 162 (1996).

\bibitem{kga_krs_01}
K. G. Aivalis, K. R. Sreenivasan, Y. Tsuji, J. C. Klewicki, and C.
A. Biltoft, Phys.\ Fluids {\bf 14}, 2439 (2002).

\bibitem{Kurien1}
S. Kurien, K. G. Aivalis, and K. R. Sreenivasan, J.\ Fluid Mech.\
{\bf 448}, 279 (2001).

\bibitem{Sreeni96}
K. R. Sreenivasan, S. I. Vainshtein, R. Bhiladvala, I. San Gil, S.
Chen, and  N. Cao, Phys.\ Rev.\ Lett.\ {\bf 77}, 1488 (1996).

\bibitem{Sreeni98}
K. R. Sreenivasan and B. Dhruva, Prog.~Theo.~Phys.\ {\bf 130}, 103
(1998).

\bibitem{Batchelor51}
G. K. Batchelor, Proc.\ Camb.\ Philos.\ Soc.\ {\bf 47}, 359
(1951).

\bibitem{stolo_sreeni1}
G. Stolovitzky and K. R. Sreenivasan, Phys.\ Rev.\ E {\bf 48}, R33
(1993).

\bibitem{stolo_sreeni2}
G. Stolovitzky, K. R. Sreenivasan, and  A. Juneja, Phys.\ Rev.\ E
{\bf 48}, R3217 (1993).

\bibitem{Fujisaka01}
H. Fujisaka and S. Grossmann, Phys.\ Rev.\ E {\bf 63}, 026305 (8
pages) (2001).

\bibitem{Arad99}
I. Arad, V. S. L'vov, and I. Procaccia,
Phys.~Rev.~E {\bf 59}, 6753 (1999).

\bibitem{Kurien2}
S. Kurien and K. R. Sreenivasan, {\em Measures of anisotropy and
the universal properties of turbulence.} In Les Houches Summer
School Proceedings 2000, pp.~53-111, Springer-EDP (2001).

\end{thebibliography}
\end{document}